\documentclass[aps,twocolumn]{revtex4}
\usepackage[dvips]{graphicx}

\begin{document}

\bibliographystyle{apsrev}

\title{Effect of quantum confinement on exciton--phonon interactions}

\author{Hui~Zhao}
\author{Sven~Wachter}
\author{Heinz~Kalt}

\affiliation{Institut f\"{u}r Angewandte Physik, Universit\"{a}t
Karlsruhe, D-76128 Karlsruhe, Germany}

\begin{abstract}
We investigate the homogeneous linewidth of localized type--I
excitons in type--II GaAs/AlAs superlattices. These localizing
centers represent the intermediate case between
quasi--two--dimensional (Q2D) and quasi--zero--dimensional
localizations. The temperature dependence of the homogeneous
linewidth is obtained with high precision from
microphotoluminescence spectra. We confirm the reduced interaction
of the excitons with their environment with decreasing
dimensionality except for the coupling to LO~phonons. The
low--temperature limit for the linewidth of these localized
excitons is five times smaller than that of Q2D excitons. The
coefficient of exciton--acoustic phonon interaction is $5\sim6$~
times smaller than that of Q2D excitons. An enhancement of the
average exciton--LO~phonon interaction by localization is found in
our sample. But this interaction is very sensitive to the detailed
structure of the localizing centers.
\end{abstract}


\maketitle

The homogeneous linewidth of exciton luminescence is one of the
most important features in excitonic dynamics in semiconductors,
since it contains directly the information about the interactions
between excitons and their environment. During the past two
decades, the homogeneous linewidth of excitons in several kinds of
quantum well and superlattice systems has been investigated
extensively in both time and frequency domains. In the time
domain, the excitonic dephasing time was measured from four--wave
mixing (FWM), and then the homogeneous linewidth could be
deduced.\cite{l571797, b349027, b406442, l681006, b592215} In the
frequency domain, the linewidth was measured directly from
photoluminescence,\cite{b335512, b4610193, b485241, jap871858}
transmission, reflection or absorption\cite{b5116785, apl611411}
and Raman spectroscopy\cite{b501792}. By modeling of experimental
data, extensive information about interactions between excitons
and acoustic phonons, LO~phonons, free carriers and other excitons
has been deduced. In these investigations, excitons are
quasi--two--dimensional (Q2D). That is, they can move freely in
the wells or are localized weakly with a localization energy of
several~meV. On the other side, the homogeneous linewidth of
quasi--zero--dimensional (Q0D) excitons confined in quantum dots,
with localization energy of several hundreds of meV, has been
studied by spatially resolved measurements.\cite{b5315743,
l744043} The comparison of these two kinds of excitons provides
information about the influence of quantum confinement on the
interactions between excitons and their environment.

In this paper, we report investigations on homogeneous linewidth
of single type--I localized excitons in GaAs/AlAs superlattices
which have a global band alignment of type II. The localization
energies of these centers are several tens of meV. Thus we can
regard these localized excitons as intermediate in dimensionality
between Q2D excitons and Q0D excitons. Furthermore, since the
investigated centers are found in a small area (1~$\mu$m in
diameter) of the same sample, we can rule out any artificial
effects which come about when comparing different samples. This
enables us to discuss the influence of localization on
exciton-phonon interactions by comparing these centers, without
disturbed by other artificial effects.

The localized excitons are studied by microphotoluminescence
($\mu$--PL). The spectral and spatial resolutions are sufficient
to detect luminescence from individual localizing centers. Our
experimental setup consists of a He flow cryostat with the sample
mounted close to a thin window. This allows the use of a
microscope objective to image (magnification 20$\times$) the
excited spot on the sample onto a pinhole. The pinhole defines the
spatial resolution. We use a 20--$\mu$m pinhole in the present
experiments, which corresponds to 1--$\mu$m detected area on the
sample's surface. The pinhole is imaged onto the entrance slit of
a 0.75--m focal length double grating spectrometer. We use a
cooled CCD to record the spectra with a spectral resolution of
30~$\mu$eV. The sample is nonresonantly and globally excited by a
He--Ne laser. The excitation intensity is about
1~$\mathrm{W/cm^{2}}$ for all of the temperature--dependent
measurements. During the measurement, the temperature of the
sample is measured with a diode temperature sensor in good thermal
contact. The temperature is stabilized by the He flow and heating
to a fluctuation of less than 0.2~K. The measurements are
performed in the range of $7\sim80$~K. We study two samples:
(i)140 periods of GaAs(3~nm)/AlAs(2.8~nm) and (ii)140 periods of
GaAs(2.3~nm)/AlAs(2.3~nm). Both samples have a type--II band
alignment, i.e., the unperturbed conduction-band minimum is in the
AlAs layer and the valence--band maximum is in the GaAs layer.
Details about the growth and the interface properties of the
samples have been reported previously.\cite{b571631} The two
samples yield quite similar results concerning the exciton--phonon
coupling. Thus we will only present data of sample (i).

\begin{figure}
 \includegraphics[width=9cm]{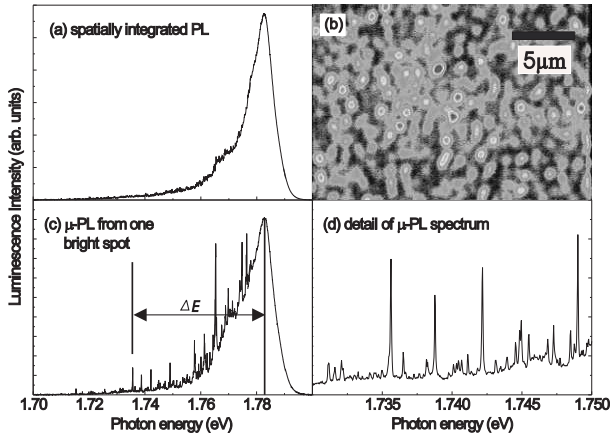}
 \caption{
 Luminescence of GaAs/AlAs superlattices at 20~K under excitation of He--Ne laser with excitation intensity of about
 1~$\mathrm{W/cm^{2}}$. (a)~Spatially integrated PL spectrum; (b)~Intensity maps; (c)~The $\mu$--PL spectrum of one
 of the bright spots; (d)~Details of (c). The intensity maps are recorded by blocking the scattered light
 from the laser, and spectrally integrating the luminescence from 1.7 to 1.8~eV.
 }
 \label{fig1}
\end{figure}

Figure~\ref{fig1} reviews the luminescence properties of the
sample at 20~K. The spatially integrated PL spectrum [Fig.~1(a)]
is composed of a zero--phonon line at about 1.782~eV and phonon
sidebands at the low--energy side. Luminescence intensity maps
[Fig.~1(b)] show an inhomogeneous distribution of the emission
intensity. We can find bright spots of about 1~$\mu$m in diameter.
This size corresponds to the resolution of the objective in our
$\mu$--PL system. The actual size of the bright spots was
determined to be about $250\sim300$~nm [full width at half maximum
(FWHM)] by scanning near--field optical microscopy with resolution
of 100~nm.

Figure~1(c) shows the $\mu$--PL spectrum of one of the bright
spots. Different from Fig.~1(a), the spectrum is dominated by
local emission from the bright spot. On the smooth background,
spectrally narrow lines are superimposed.\cite{b5915862}
Figure~1(d) provides a closer sight of these narrow lines. We have
found that the narrow lines observed in Fig.~1(c) could be divided
into two groups according to their different temperature
behaviors.\cite{b5915862} For the lines on the low--energy side of
the zero--phonon line, i.e., $1.75\sim1.782$~eV, the spectral
weight shifts red with rising temperature, and their integrated
intensity drops. These lines stem from localized type--II states.
For the lines in the spectral range of the AlAs LO--phonon
replica, i.e., below 1.74~eV, the spectral weight does not change
significantly, and their intensities increase exponentially with
temperature up to 50~K. We have proved that, although the global
band alignment of this sample is type II, the layer thickness
fluctuations give rise to local changes in the band alignment
toward type I. Recombination of excitons localized in these
type--I centers is the origin of the narrow lines in the energy
range of $1.69\sim1.74$~eV. The population mechanism of these
localized states has been proved to be electron tunneling from
AlAs layers to GaAs layers.\cite{b5915862} Since some of these
narrow lines are well separated in energy, we can resolve each of
them without serious disturbance by adjacent lines. Thus we can
analyze the luminescence from single localizing centers in the
GaAs layers.

\begin{figure}
 \includegraphics[width=9cm]{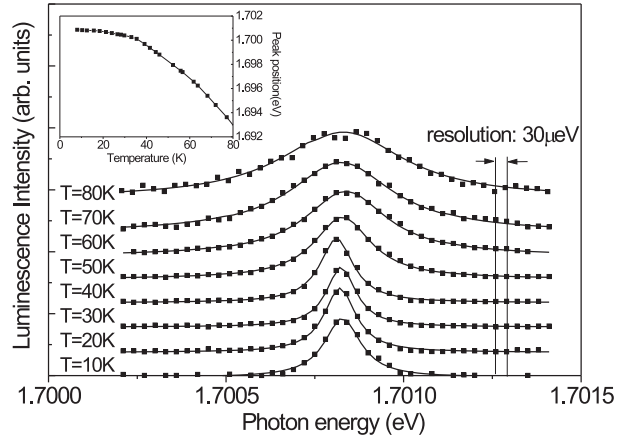}
 \caption{
 An example of the measured narrow lines (squares) and the corresponding Lorentzian fit (solid line).
 The peak positions have been shifted for better illumination. For the actual positions under different temperatures, see the inset.
 }
 \label{fig2}
\end{figure}

In order to investigate the temperature dependence of the
linewidth of excitons localized in the type--I centers, $\mu$--PL
spectra from several bright spots were measured in the temperature
range of $7\sim80$~K. To check the possible spectral wandering
during the integration, we measured the spectra with different
integration times. The spectral shape and the linewidth keep
unchanged as we vary the integration time in the range of
$50~\mathrm{ms}\sim30~\mathrm{s}$. We also measured the sequence
of the spectra with an integration time of 50~ms and an
interruption time of 1~s. We didn't find any change of the peak
position among these spectra. Thus, the spectral wandering of the
sample can be neglected. The narrow lines were fitted by
Lorentzian line shapes to obtain the linewidth (FWHM) in each
temperature. An example of the narrow lines and the fitting curves
is shown in Fig.~\ref{fig2}. Recently, Besombes {\it et
al}.\cite{b63155307} found that the line shape of the luminescence
from strongly confined CdTe quantum dot deviates from Lorentzian
shape with increasing the temperature. The whole spectrum is
composed of a zero--phonon line and an additional acoustic--phonon
sideband which results from lattice relaxation due to
exciton-phonon coupled states. In InAs/GaAs system, both
Lorentzian\cite{b65033313} and the non-Lorentzian\cite{l87157401}
line shapes have been observed recently. For GaAs quantum dots,
the PL spectrum has been shown to be of Lorentzian line
shape.\cite{l763005} Also in our experiments on centers of
intermediate confinement in GaAs/AlAs superlattices, we find no
indications of additional phonon sidebands. The spectral line
shape can be well fitted by Lorentzian function up to 80~K (see
Fig.~2).

\begin{figure}
 \includegraphics[width=9cm]{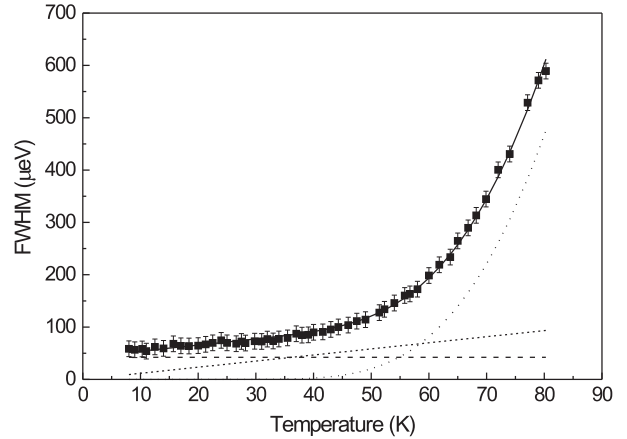}
 \caption{
 Temperature dependence of the homogeneous linewidth of one narrow line. The experimental data (squares)
 were fitted by Eq.~(1) (solid line). The contributions to the linewidth from acoustic--phonon scattering and LO--phonon
 scattering are also shown (short--dashed and dotted lines, respectively). The dashed line represents the low--temperature limit of the linewidth.
 }
 \label{fig3}
\end{figure}

In general, the temperature dependent homogeneous linewidth of the
exciton resonance is written as (see, e.g.,
Ref.~\onlinecite{b4211218})
\begin{equation} \label{eq:eq1}
\Gamma_{homo}(T)=\Gamma_{0}+\gamma_{AC}T+\gamma_{LO}[\mathrm{exp}(\hbar\omega_{LO}/k_{B}T)-1]^{-1}
\end{equation}
where the term linear in temperature is due to exciton scattering
with acoustic phonons, and the term nonlinear in temperature is
due to interactions with LO~phonons. The coefficients
$\gamma_{AC}$ and $\gamma_{LO}$ represent the strength of the
exciton--acoustic--phonon interaction and exciton--LO--phonon
interaction, respectively. The first term in Eq.~(1) is the
low--temperature limit of the linewidth. The temperature
dependence of the linewidth deduced from the Lorentzian fits was
then fitted by Eq.~(1), for several well--separated narrow lines.
We show one of the fitting results in Fig.~\ref{fig3}, as an
example. The contributions to the linewidth from acoustic--phonon
scattering and LO--phonon scattering are also shown in this figure
(short--dashed and dotted lines, respectively). The dashed line
represents low--temperature limit of the linewidth. For this
narrow line, the fitting parameters are $\Gamma_{0}$ = 42 $\mu$eV,
$\gamma_{AC}$ = 1.2 $\mu$eV/K, and $\gamma_{LO}$= 85.7 meV. Since
the fitting contains three free parameters, it is necessary to
check the sensitivity of these parameters on fitting process. For
this purpose, we vary each of these parameters from its optimal
value, to check the variation of mean--square deviation between
fitting curve and experimental data. We find that a 10~\%
deviation of $\Gamma_{0}$, $\gamma_{AC}$, or $\gamma_{LO}$ from
their optimal values corresponds increase of 60~\%, 94~\% or
580~\% in the mean--square deviation, respectively. This result
insures the safety of the fitting process. Another possible
problem in extracting the parameters from the fitting is whether
the temperature range ($7\sim80$~K) is large enough for an
accurate determination of the parameters, especially for the
$\gamma_{LO}$. In the investigations of the Q2D excitons, the
linewidths were measured up to 150~K (Ref.~\onlinecite{b335512})
or even room temperature\cite{l681006, jap871858, b5116785}. In
the present study, however, the luminescence of these localized
excitons at a temperature above 80~K is too weak for an accurate
determination of the linewidth. To check the possible errors
introduced by this relatively small temperature range, we redraw
the results obtained by the larger temperature range
measurements,\cite{b335512,l681006,jap871858,b5116785} and fit by
Eq.~(1) all the data in the whole temperature range or only the
data obtained below 80~K, respectively. We find that for all of
the results checked, the difference of the parameter values
obtained from the two fits using different ranges of data is less
than 10~\% for $\gamma_{LO}$ and less than 3~\% for $\gamma_{AC}$.
This proves the temperature range of $7\sim80$~K is enough for the
determination of the linewidth--temperature curve with a
satisfactory accuracy. Furthermore, the data density in the
present study (about 50 data points in the temperature range of
$7\sim80$~K) is high enough for an accurate fitting.

\begin{figure}
 \includegraphics[width=9cm]{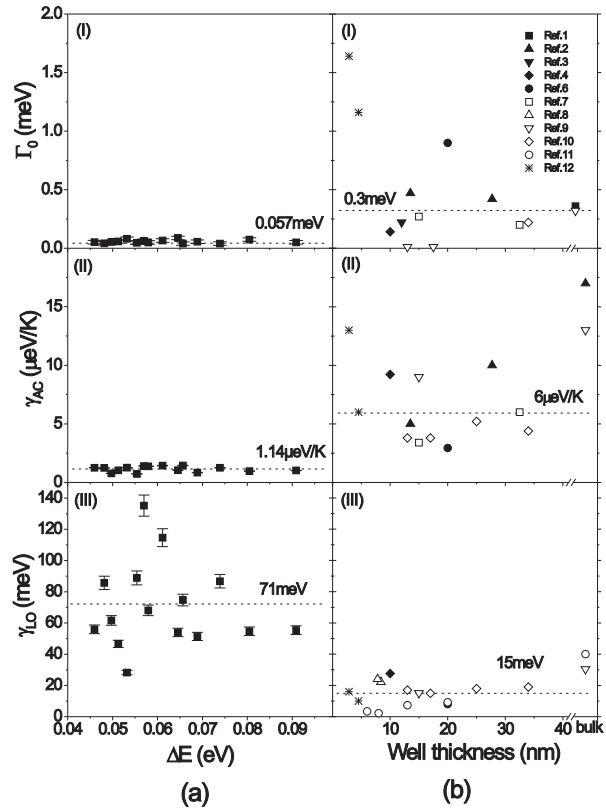}
 \caption{
  (a) Low--temperature limits of linewidth (I), exciton--acoustic--phonon interaction coefficients (II) and exciton--LO--phonon
  interaction coefficients (III) gained by fitting the measured $\Gamma \sim T$ relations (see Fig.~3).
  The error bars show the uncertainty of the fitting due to a limited temperature range, as discussed in the text.
  The dashed lines represent the average values;
  (b) The corresponding available values for quasi--two--dimensional excitons in GaAs quantum wells and excitons in
  GaAs bulk and superlattices. References for all data points are
  listed in (b)I.
 }
 \label{fig4}
\end{figure}

In Figure 4(a), we list the fitting results of $\Gamma_{0}$,
$\gamma_{AC}$, and $\gamma_{LO}$ of the narrow lines analyzed in
the present study. In order to distinguish the narrow lines, we
define $\Delta E$ as the energy difference between the
corresponding narrow line and the peak of zero--phonon line. We
show an example of this definition for one narrow line in
Fig.~1(c). Such a quantity is temperature independent. We note
that $\Delta E$ is close to, but not exactly, the localization
energy of the corresponding center, since we cannot regard the
zero--phonon--line as the mobility edge exactly. However, such a
difference will not influence our discussions. For comparison, we
list in Fig.~4(b) the available data of GaAs Q2D excitons deduced
from FWM, photoluminescence, or other methods, as a function of
the thickness of the GaAs layers. The results of GaAs bulk and
superlattices are also listed in this figure, but not included in
the calculations of average values, which are shown as dashed
lines in Fig.~4(b). In the viewpoint of quantum confinement, we
regard the localized excitons investigated here as the
intermediate case between Q2D excitons in quantum wells and Q0D
excitons in quantum dots. In the following, we will compare the
$\Gamma_{0}$, $\gamma_{AC}$, and $\gamma_{LO}$ of localized
excitons obtained in the present study with that of the other two
cases, to discuss the influence of confinement on exciton--phonon
interactions in semiconductors.

At first, we discuss the low--temperature limit of the linewidth.
We obtain the average value of $\Gamma_{0}$ to be $0.057~(\pm
0.014)$ meV for these localized excitons [dashed line in
Fig.~4(a)I]. This value is five times smaller than the average
value of Q2D excitons [dashed line in Fig.~4(b)I]. That indicates
that the additional in--plane confinement in localizing centers
reduces $\Gamma_{0}$. In order to investigate the contribution of
intercarrier scattering to this linewidth, we measured the
excitation intensity dependence of the linewidth at 7~K. In the
intensity range of $1\sim10~\mathrm{W/cm^{2}}$, the linewidth
keeps unchanged, while in the range of
$10\sim5000~\mathrm{W/cm^{2}}$ the linewidth increases slowly with
a slope of $0.01\sim0.02$~$\mu$eV/($\mathrm{W/cm^{2}}$). Due to
the complicated population mechanism of these localizing centers
(electron tunneling from AlAs layers to GaAs layers), we are not
able to relate the excitation intensity to the actual carrier
density in the sample. However, we can conclude that the
excitation intensity used in the temperature--dependent
measurement is quite low, and the intercarrier interaction can be
neglected. The intrinsic lifetimes of excitons in quantum wires
and quantum dots have been calculated to be several hundreds of
picosecounds.\cite{l693393} However, recent experiment reveals a
16--ps intrinsic lifetime of excitons in GaAs quantum
dots.\cite{s2932224} The linewidth obtained in the present study
corresponds to a lifetime of 22~ps, which is consistent with this
new finding.

Second, we discuss the influence of confinement on
exciton--acoustic--phonon interaction. It has been found
before\cite{b349027, b501792} and confirmed again
recently\cite{jap871858} that $\gamma_{AC}$ of Q2D excitons is
smaller than that in bulk GaAs [Fig.~4(b)II] . The average value
of $\gamma_{AC}$ of the localized excitons obtained in the present
investigation is $1.14~(\pm 0.24)$~$\mu$eV/K, about $5\sim6$~times
smaller than that of Q2D excitons [dashed lines in Figs.~4(a)II
and 4(b)II]. The result suggests a reduction of
exciton--acoustic--phonon interaction by localization. In quantum
wires, such a reduction has been found by direct comparison of
free and localized excitons in FWM measurements.\cite{b492993}
Furthermore, in quantum dots, the homogeneous linewidth has been
found to be almost constant up to 50~K.\cite{b5315743, l744043}
These results suggest the extremely small $\gamma_{AC}$ for Q0D
excitons. The weaker acoustic--phonon interaction with Q0D
excitons than with Q2D excitons has also been confirmed in II--VI
systems (see, for example, Ref.~\onlinecite{b63155307}). The whole
evolution discussed above, from bulk via Q2D excitons to localized
excitons (this study) and to Q0D excitons, implies strongly that
the interaction between exciton and acoustic phonon is steadily
reduced by increasing confinement. Such a behavior is consistent
with previous theoretical predictions.\cite{jlumi44347, b428947}
In a confined system, the final state of phonon scattering is not
always available due to the discrete energy level scheme. Thus, by
increasing the quantum confinement, the appearance of the discrete
energy levels induces a decrease of the acoustic phonon
interaction. We note that when the confinement is so strong that
the energy level space is larger then the thermal energy $k_{B}$T,
a further increase of the confinement does not further reduce the
interaction, since the level space has already been large enough
for this bottleneck effect. In this regime, additional effects
like lattice relaxation\cite{b63155307} can influence the
dependence of the acoustic phonon interaction on the quantum dot
size. Theoretical calculations revealed an increase, rather than
decrease, of the acoustic--phonon coupling when further reducing
the size of the quantum dots in this
regime.\cite{b602638,b63155307}

The parameters $\Gamma_{0}$ and $\gamma_{AC}$ of the localized
excitons obtained here are almost independent of $\Delta E$ in the
range of $0.04\sim0.09$~eV [Figs.~4(a)I and 4(a)II]. But, for
$\gamma_{LO}$, we find a totally different behavior in the same
energy range. The values of $\gamma_{LO}$ vary in the range of
$30\sim140$~meV, with no obvious systematic dependence on $\Delta
E$. The fluctuations of $\Gamma_{0}$ and $\gamma_{AC}$, which are
also obtained in the same fitting process, are all less than
25~\%. We attribute the observed scattering in the homogeneous
linewidth to an intrinsic feature of the exciton--LO--phonon
coupling. In localizing centers, the energy level scheme of
excitons is determined by the detailed structure of the center.
Due to the monochromatic feature of the LO--phonon dispersion, the
exciton--LO--phonon scattering rate depends sensitively on the
level scheme. For the center in which the energy level scheme
matches the LO--phonon energy well, a strong coupling is observed.
In contrast to the LO phonons, the dispersion of the acoustic
phonons distributes over a relative wide energy range. Thus, the
exciton--acoustic--phonon scattering rate is less sensitive to the
detailed structure of the localizing centers. In fact, we do not
find the pronounced resonant behavior for the acoustic--phonon
coupling [see Fig.~4(a)II].

In the strongly confined quantum dots, the explicit size and shape
of the localizing potential determines the spatial extension and
anisotropy of the electron--hole wave function as well as the
electron--hole overlap. This has significant influence on the
exciton--phonon interaction.\cite{l834654,b64241305} In strongly
confined CdTe quantum dots, a mixing of the exciton and
acoustic--phonon modes, which cannot be described by perturbation
treatment, has been proposed.\cite{b63155307} That is, the exciton
locally distorts the lattice of the dot. This lattice distortion
is important for small quantum dots which sizes are comparable
with the exciton Bohr radius. For example in II--VI and InAs/GaAs
systems, an induced non-Lorentzian broadening have been
observed.\cite{b63155307,l87157401} However, the localizing
centers studied here are much larger than the Bohr radius. Thus
the distortion is less important, and we do not observe strong
deviations from a Lorentzian lineshape even at a temperature of
80~K. For the same reason, the influence of the potential size and
shape on the electron--hole wave function is also less pronounced
than that in strongly confined quantum dots. So we observe only
small variations in the acoustic--phonon coupling strength among
these localizing centers with different sizes and shapes.

Despite of the scattering behavior, we can still deduce the
enhancement of the exciton--LO--phonon interaction in localized
excitons with respect to Q2D excitons. The average value of
$\gamma_{LO}$ is 71~meV, about five times larger than that of Q2D
excitons [Figs.~4(a)III and 4(b)III]. The enhancement of
exciton--LO--phonon interaction by localization induced by
alloying fluctuations in alloy $\mathrm{GaAs_{1-x}P_{x}}$ has been
found by resonant Raman spectroscopy.\cite{b547921} A similar
enhancement was also found in GaN quantum wells.\cite{apl702882}
In those investigations, the LO--phonon replica was used to detect
the exciton--phonon interaction. In the present study, we detected
luminescence from excitons localized by thickness fluctuations by
$\mu$-PL. The agreements between different experimental methods as
well as the different origins of localization confirm that the
additional in--plane confinement on excitons enhances the
exciton--LO--phonon interaction.

Up to now, the exciton--LO phonon interaction in quantum dots is
still an open problem. In CdTe quantum dots, Besombes {\it et
al}.\cite{b63155307} found that the exciton--LO--phonon scattering
is not efficient up to 60~K, while Heitz and
co--workers\cite{l834654, apl773746} observed enhanced
exciton--LO--phonon interaction in InAs/GaAs self--organized
quantum dots by measuring the phonon--assisted exciton
transitions. The Huang--Rhys parameter was found to be five times
larger than in bulk InAs. The enhancement was attributed to the
quantum confinement and piezoelectric effect. Our result confirms
qualitatively the latter finding. According to the extremely
sensitive dependence of exciton--LO--phonon interaction on the
detailed structures of localizing centers, we suggest that much
care should be taken when comparing experimental results of this
interaction in different quantum dot samples, since the detailed
structure of the dots can be totally different, and this may
influence the strength of the interaction to a great extent.

In summary, we have measured the homogeneous linewidth of type--I
localized excitons in type--II GaAs/AlAs superlattices using
$\mu$--PL. These excitons, with a localization energy of several
tens of meV, can be regarded as intermediate case between Q2D
excitons (free or weakly localized excitons in quantum wells) and
Q0D excitons in quantum dots with confinement energy of several
hundred meV. The low--temperature limit of the linewidth,
$\Gamma_{0}$, of these localized excitons is found to be five
times smaller than that of Q2D excitons. We obtain a
$5\sim6$~times smaller exciton--acoustic--phonon interaction
coefficient, $\gamma_{AC}$, for the localized excitons with
respect to that of Q2D excitons. Together with a comparison of
exciton data in bulk and quantum dots, the reduction of
exciton--acoustic--phonon interaction by confinement is confirmed.
In contrast to the results on $\Gamma_{0}$ and $\gamma_{AC}$,
which are independent of localization energy, the coupling to
LO~phonons, $\gamma_{LO}$, shows strong variations. This finding
is attributed to the strong influence of the energy level scheme
on the exciton--LO--phonon coupling. In average, we confirm an
enhancement of exciton--LO--phonon interaction by
localization.\newline

We gratefully acknowledge the growth of the excellent samples by
W.~Braun and K.~Ploog, and helpful discussions with K.~Kheng. This
work was supported by the Deutsche Forschungsgemeinschaft.

\end{document}